\documentclass[preprint,showpacs,preprintnumbers,amsmath,amssymb]{revtex4-1}

\usepackage{graphicx}
\usepackage{dcolumn}

\usepackage{bm}
\usepackage{bm}
\usepackage{float}
\usepackage{color}
\usepackage[T1]{fontenc}       

\usepackage{notes2bib}

\begin{document}
\date{\today}

\renewcommand{\refname}{}

\newcommand{\bfr}{\mathbf{r}}
\newcommand{\bfR}{\mathbf{R}}
\newcommand{\bfk}{\mathbf{k}}
\newcommand{\bfK}{\mathbf{K}}
\newcommand{\bfq}{\mathbf{q}}
\newcommand{\bfg}{\mathbf{g}}
\newcommand{\bfG}{\mathbf{G}}
\newcommand{\bfa}{\mathbf{a}}
\newcommand{\bfb}{\mathbf{b}}
\newcommand{\bfH}{\mathbf{H}}
\newcommand{\bfV}{\mathbf{V}}
\newcommand{\bfI}{\mathbf{I}}
\newcommand{\bfE}{\mathbf{E}}
\newcommand{\bfx}{\mathbf{x}}
\newcommand{\etal}{\textit{et~al.}}



\title{Impact of single and double oxygen vacancies on electronic transport in Fe/MgO/Fe  magnetic tunnel junctions}

\author{B. Taudul}
\affiliation{Universit\'{e} de Strasbourg, Institut de Physique et Chimie des Mat\'{e}riaux de Strasbourg, Unistra-CNRS UMR 7504, 23 Rue du Loess, BP 43, 67034 Strasbourg Cedex 2, France}
\author{ M. Bowen}
\affiliation{Universit\'{e} de Strasbourg, Institut de Physique et Chimie des Mat\'{e}riaux de Strasbourg, Unistra-CNRS UMR 7504, 23 Rue du Loess, BP 43, 67034 Strasbourg Cedex 2, France}
\author{ M. Alouani}
\affiliation{Universit\'{e} de Strasbourg, Institut de Physique et Chimie des Mat\'{e}riaux de Strasbourg, Unistra-CNRS UMR 7504, 23 Rue du Loess, BP 43, 67034 Strasbourg Cedex 2, France}

\begin{abstract}

The combination of a low tunnelling barrier height and a large tunnelling
    magnetoresistance (TMR) ratio in MgO-class magnetic tunnel junctions has
    enabled next-generation information storage and bio-inspired computing
    solutions thanks to the spin transfer torque effect. Recent literature has
    proposed that this synergistic combination arises from the electronic
    properties of oxygen vacancies. To explicitly understand their impact on
    spin-polarized transport, we have computed the electronic and transport
    properties of single (F centers) and paired (M centers) oxygen vacancies
    using density functional theory and the projector augmented wave (PAW)
    method. These point defects can generate barrier heights as low as 0.4V for
    FeCo electrodes irrespective of the defect's spatial position within the
    barrier, and of the orientation of the M center. These defects promote a
    strong decrease in the conductance of the spin up channel in the MTJ's
    parallel (P) magnetic state that mainly accounts for an order-of-magnitude drop
    in TMR, from $\approx$10000\%  in the ideal case toward values more in line with experiment. When placed in the
    middle layer of the MgO barrier,  the F center introduces additional P
    $\uparrow$ transmission away from the $\Gamma$ point. 
    This scattering lowers TMR to 145\%. In contrast, the M center merely
    broadens this transmission around $\Gamma$, thereby boosting TMR to \%315. 
    Rotating a M center so as to partly point along the transmission direction sharpens transmission around $\Gamma$, 
    further increasing TMR to 1423\%. When these defects are
    placed at the MTJ interface, the transmission and ensuing TMR, 
    which reaches $\approx$4000\%, suggest that such junctions behave as
    would an ideal MTJ, only with a much lower barrier height. Our results thus
    theoretically reconcile the concurrent observations of high TMR and low
    barrier heights, in line with experimental preparation techniques such as
    post-deposition oxidation of metallic Mg, which can generate oxygen
    vacancies at the lower MTJ interface, and annealing which can promote M
    centers over F centers. Our theory is also in line with an origin of
    perpendicular magnetic anisotropy in terms of oxygen vacancies at MTJ
    interfaces. The effective size of these vacancies sets a limit for
    both the barrier thickness, in line with experiment, as well as for the MTJ's lateral dimension. 
    Our work provides a much-needed theoretical basis to move beyond the mostly unsuspected, 
    fortuitous defect engineering of spintronic performance that has thus far propelled MgO-based spintronics and its applications.
\end{abstract}

\pacs{71.15.Mb, 73.20.-r,73.40.-c,73.40.Qv}

\maketitle


\section{Introduction}

Spintronic research exploits both charge and  spin degrees of freedom in
solid-state systems\cite{Zutic2004,Makarov2016,Lu2016,Peng2014}.  A widely
studied spintronic device is the magnetic tunnel junction (MTJ), composed of
two ferromagnetic metallic electrodes separated by an ultrathin dielectric. The
electrical resistance of the MTJ depends on the relative orientation of the
electrode magnetizations, which can be controlled by an external magnetic field
or a spin-polarized current\cite{NatureMaterials_2013_Locatelli_Grollier_Spin-torque}.  This change in resistance
is called \textit{tunnelling magnetoresistance} (TMR) and is defined as

\begin{equation}
TMR =\frac{R_{\rm AP} - R_{\rm P}}{R_{\rm P}},
\end{equation}
where $R_{\rm P}$ and $R_{\rm AP}$ are the resistances for parallel (P) and antiparallel (AP) configurations of the two magnetizations of the electrodes.

Initial MTJs with an Al$_{2}$O$_{3}$ tunnel barrier exhibited a maximum TMR of
70\% at room temperature.\cite{Moodera1995,Wang2004}  Theoretical studies of
magnetotransport across Fe/ZnSe/Fe\cite{MacLaren_PRB59_1999} and
Fe/MgO/Fe\cite{MacLaren_PRB59_1999,Butler2005,Butler_2008,Zhang2003,Butler_lateral_variation_PRB63_2001}
MTJs revealed that the TMR can be greatly increased if the amorphous barrier is
replaced by a crystalline one, such that certain orbitals with a high spin
polarization in the electrodes preferentially tunnel across the barrier.
Nowadays, textured FeCoB/MgO/FeCoB MTJs with TMR values above 600\% at room
temperature \cite{ikeda_tunnel_2008} offer promising prospects for data
read-out, storage and processing, magnetic
sensors\cite{kent_new_2015,Nature,ikeda}.

Despite the importance of these MTJ technologies, an understanding of exactly
how the device operates remains a work in progress. Indeed, the TMR effect is a
complex phenomenon that depends strongly on the electronic structure of the
electrodes, the properties of the insulating barrier and on the chemical
bonding at the MTJ interface. As an illustration, consider how the success of
achieving high TMR concurrently with low barrier heights
required\cite{Halisdemir2016} to implement spin transfer torque toward
these MTJ technologies\cite{kent_new_2015,Nature,ikeda,NatureMaterials_2013_Locatelli_Grollier_Spin-torque} imply
that structural defects, which may lower TMR from the 10000\% theoretical
prediction, may actually play a beneficial spintronic role here.

Several causes for an effective deviation from the MTJ's ideal structure have
been considered. Experiments often reveal the presence of interface oxidation,
which alters the nature of chemical bonding at the interface between the
ferromagnetic electrodes and the MgO spacer and can degrade
TMR\cite{Meyerheim2001GeometricalAC}. The combination of
theoretical\cite{Zhang_oxidation_2003} and
experimental\cite{Bonell_oxidation_exp_2009} studies showed that, even if one
includes the interface disorder or the oxidation of interfacial Fe layer, the
drastic drop of TMR cannot be fully explained by this mechanism alone.

As another cause, atomic diffusion may occur during the sample preparation and
annealing\cite{App_Phys_Lett_89_232510_2006}. In particular, boron
diffusion into the MgO barrier (forming boron oxides), or its segregation at
the CoFe/MgO interface, has been
studied\cite{Miyajima_B_to_capping_layer_APL_2009,Kozina_B_diffusion_in_Ta_APL_2010,Pinitsoontorn_B_segregation_at_interface_APL_2008,Rumaiz_B_at_interface_and_MgO_APL_2011},
but this is not always the case
\cite{Kurt_No_B_in_MgO_APL_2010,Mukherjee_control_Boron_PRB91_2015,Wang_COFeBMgO_structure_nanolett_2016}.
Rather, at a proper annealing temperature, boron does not diffuse into MgO but
rather goes further away from the interfaces.  Even if boron diffuses into MgO,
it was shown theoretically that this should not create additional states within
the MgO band gap \cite{Bai_B_induced_symm_reduction_theory_PRB87_2013}.

Finally, another source can be imperfections in the MgO spacer itself, such as
grain boundaries and point defects. The impact of the grain boundaries on the
electronic structure and on the transport is difficult to address. Nonetheless,
it was shown by Mizuguchi \etal \cite{Mizuguchi_GB-exp_2007} that the
tunnelling current flows uniformly despite the existence of the grain
boundaries and hence the device performance is not affected considerably by
this kinds of defects.  Moreover, the combined experimental and theoretical
investigations of Bean \etal \cite{Bean_GB_2017} showed that grain boundaries
can cause a decrease of the effective barrier of MgO but this band gap decrease
can not explain the observed low barrier heights\cite{SchleicherNC}.

Point defects, on the other hand, can promote localized states within the band gap of
MgO, giving rise to a variety of interesting optical, catalytic and transport
properties that are absent in the ideal crystalline
material\cite{Rosenblat1989}.  The most plausible imperfections are oxygen and
magnesium vacancies, denoted F and V centers respectively. They can appear in a
neutral, singly charged or doubly charged state that is denoted as F$^{+}$,
F$^{2+}$, V$^{-}$ V$^{2-}$, respectively. Moreover, two point defects can form
a paired vacancy: two F centers form a F$_2$ pair of oxygen vacancies, which is
a M center when they are nearest-neighbor on the oxygen sublattice. A F center
can also combine with a V center to form a MgO vacancy.

As discussed by Gibson \etal\cite{vacanciesMgO}, oxygen vacancies exhibit the
lowest formation energy, which implies that this species of defects is more
likely to occur in MgO. This defect species promotes localized states in the
band gap of MgO and can affect the optical and the electrical properties of the
dielectric \cite{vacanciesMgO,velevBe1,velev2,SchleicherNC}.  As a result, the
barrier heights encountered by the propagating electrons are locally
reduced\cite{kim_control_2010,Studniarek2016}. The electrons can then
tunnel through the barrier via these additional states with different
scattering rates than those for an ideal barrier. This would explain
experimental reports of a barrier height in MgO MTJs that is far below the
nominal value of 3.9 eV (see Tab.~\ref{tab:bbarriers_exp}).

\begin{table}[h!]
\vspace{0.2cm}
\begin{tabular}{l  l  l}
MTJ & TMR ($\%$) &  Barrier height (eV)    \\
\hline
\hline
    Fe/MgO/Fe &  130 (190$_{1K}$)  & 0.38/0.82\cite{miao_disturbance_2008} \\
Fe/MgO/Fe &  180 (247$_{20K}$) &  0.39\cite{NatureMaterials_2004_Yuasa_Ando_Giant}  \\
FeCo/MgO/FeCo & 120-220  &  1.1-1.7 \cite{Stuart_FeCoMgO_TMR_220_2004} \\
FeCoB/MgO/FeCoB & 100    &  0.62/0.5\cite{SchleicherNC} \\
Fe/MgO/FeCo &  23$_{4.2K}$/ 20$_{70K}$&   0.9\cite{Mitani_09barreir_FeMgOFeCo_2003} \\
\hline
\hline
\end{tabular}
\caption{Experimental TMR and barrier heights for MTJs based on MgO.}
\label{tab:bbarriers_exp}
\end{table}

Although oxygen vacancies within MgO appear to play an important role not only
toward MTJ performance, but also spin transfer torque\cite{Halisdemir2016},
their clear
identification\cite{lu_spin-polarized_2009,Teixeira}
and impact on the tunnelling current has remained a work in progress. According
to theory,\cite{velev2,velevBe1,Ke2010} single oxygen vacancies should create
barrier heights of about 1.1~eV for the tunnelling electrons and decrease the
resulting TMR. Even if we consider more F-type vacancies within
MgO\cite{Ke2010}, the general conclusion is that these vacancies should degrade
the TMR ratio\cite{miao_disturbance_2008}.  On the other hand, in the
presence of a 0.4eV barrier height, coherent transport seem to be spintronically
favorable\cite{NatureMaterials_2004_Yuasa_Ando_Giant,SchleicherNC,mckenna}.  Indeed, it was suggested by Schleicher
\textit{et al.} \cite{SchleicherNC} that this barrier arise from M centers,
which preserve coherent transport according to McKenna and
Blumberg\cite{mckenna}. Only recently has a unified experimental/theoretical
picture of the potential landscape due to F and M centers
emerged\cite{Advanced_Materials_2017,SchleicherFeBpaper}.

In this work, we study the electronic properties of F and M centers in MgO, and
their impact on spin and symmetry polarized transport in Fe/MgO/Fe junctions
employing density functional theory. The transport is calculated within the
Landauer-B\"uttiker formalism  as implemented in PW\textsc{cond}
\cite{PWcond_2}. We show that the position of defect levels with respect to the
Fermi level of the MTJs is robust against the type of exchange and correlation
functional, and do not depend on the defect's spatial position in the barrier.
Our results indicate that M centers can account for the experimental barrier
height of 0.4~eV, and promote improved TMR relative to F centers. We also find
that the defect's position within the barrier thickness strongly conditions
spintronic performance. Relative to an ideal MTJ, this performance is mostly
unaffected when the F or M center is close to an interface, but is reduced when
the vacancy is moved onto the barrier's middle monolayer. In that case,
rotating the M center restores stronger TMR. When judicious, we discuss how our
theoretical framework of spintronic tunnelling across MgO in the presence of F
and M centers adheres to experiment. The spatial extent of the F and M center
suggest that a MTJ with a lateral size $\approx$ 2nm can still exhibit high
TMR\cite{Advanced_Materials_2017}.

The paper is organized as follows: In section II we present the details of our
calculations and methodology. In sec. III we compute and discuss the electronic
ground state properties of F and M centers either in bulk MgO or incorporated
into  MTJs. We compare the defect level positions obtained theoretically with
the experimental data and discuss its change with respect to the types of
electrodes. In sec. IV we show the results of transmission calculations for
Fe/MgO/Fe MTJ with oxygen vacancies generated within MgO spacer. We explain the
importance of the geometrical position of the vacancy with respect to the
interface and the orientation of the defect plane for the M center. In the last
section, we conclude the paper with a general discussion.

\section{Methodology}

\subsection{Ground state calculations}

To calculate the electronic structure of MgO with F and M centers, denoted as
F(M)-MgO, we created the F/M centers by removing one/two neutral oxygen atoms
from a simple cubic supercell containing 64 atoms. We used the experimental MgO
lattice constant of $a_{{\rm MgO}}=4.21$~\AA.  These calculations were
performed using VASP package \cite{VASP,PAWVASP} based on the projector augmented
wave (PAW) method\cite{Blochl_PAW_1994} and the Pedrew, Burke, Enzerhof (PBE)
generalized gradient approximation\cite{PBE_1996} for the  exchange-correlation
potential. The kinetic energy cutoff value of 500 eV for the plane wave basis
set and  the convergence criterion for the total  energy of 1$\mu$eV is used.
The structures with defects were relaxed by requiring that the forces acting on
atoms be less than 0.001 eV/\AA.  Due to the large size of the supercell, we
found that a $k$-point mesh of 4$\times$4$\times$4 using the Methfessel-Paxton
method with a smearing of $\tau=0.2$~eV yields a satisfactory convergence of
localized states resulting from these defects.  Since it is well known that the
generalized gradient approximation (GGA) underestimates the size of the band
gap, we also used the hybrid Heyd, Scuseria, Ernzerhof (HSE)
functional\cite{HSE_solids_2006} to correct the band gap and verify whether the
defect level positions with respect to the Fermi level depends on the type of 
functional employed.  The HSE hybrid functional mixes a portion of exact Fock
exchange with that of DFT using an adjustable parameter $\mu$.  We found that,
by increasing the Fock exchange to 43\% in HSE06, we can reproduce the
experimental band gap of MgO. This parameter is then used to calculate the
defect levels in the MgO MTJs employing HSE06.

\begin{figure}[!ht]
\includegraphics[width=\linewidth]{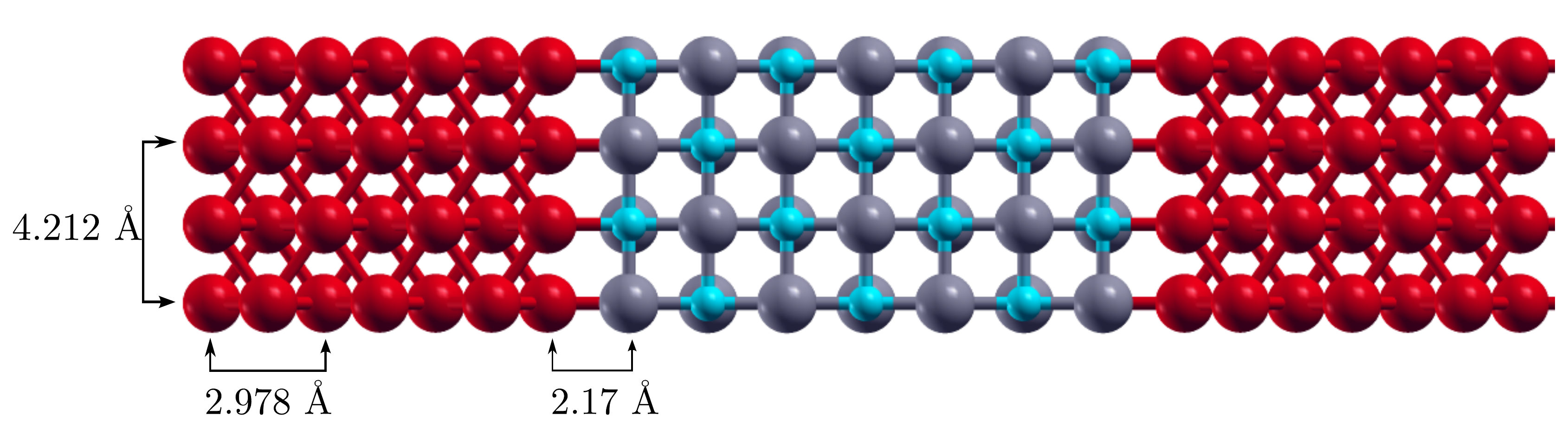}
\caption{Schematic representation of Fe/MgO junction with lattice parameters as
indicated (Red atoms are Fe, blue O and grey Mg).} \label{fgr:Fe_MgO_junction}
\end{figure}

To determine the positions of the defect levels with respect to the Fermi
energy and compare them with experiment, we used more realistic Fe/MgO/Fe MTJs.
At the metal/insulator interface, the electronic transfer between the two
materials and the metal induced gap states (MIGS) in the band gap of MgO will
peg the Fermi level position for the junction and establish the energy position
of the defect levels accordingly.  The geometry of the MgO/Fe supercell was
based on the experimental results: the Fe conventional unit cell was rotated
by a 45$^\circ$ with respect to that of MgO to match the lattice constants of
both materials and avoid strains in the structure: $a_{{\rm MgO}}=\sqrt{2}
a_{{\rm Fe}}$.  In addition, oxygen atoms were placed on top of Fe atoms and
the Fe-O distance at the interface was fixed to 2.17 \AA~ following previous
theoretical predictions\cite{Butler,velevBe1}. It is important to notice that
the measured Fe-O distance is in the range of 2~\AA
\cite{Urano1988,Wulfhekel_LEED_2001}  to 2.2~\AA
\cite{Meyerheim2001GeometricalAC}.  We fixed the lattice constant of MgO
($a_{\rm{MgO}}$ = 4.21~\AA ) and adjusted the lateral lattice parameter of the
electrodes  to it.  This choice reflects the experimental
evidence\cite{Japanesejournalappliedphysics_2005_Hayakawa_Ohno_Dependence} that
the annealing of FeCoB/MgO-based MTJs led to a recrystallization of the
electrode/barrier interfaces so as to adopt the MgO lattice constant. The
lattice parameter along the $z$ axis was rescaled accordingly. The structure of
the junction and relevant parameters are indicated in
Fig.~\ref{fgr:Fe_MgO_junction}.

The experimental FeCoB electrodes are initially amorphous alloys whose
interface with MgO adopts the latter's (001) texture upon annealing\cite{Japanesejournalappliedphysics_2005_Hayakawa_Ohno_Dependence}. However,
the exact arrangement of Fe and Co atoms is not really known. Moreover, the
role of boron or its influence on the electronic structure of FeCo electrode is
not clear. Therefore in our work we decided to consider a simplified scenario
in which the FeCo electrode has the same structure as that of Fe. For the
FeCo/MgO calculations, we have alternated the Fe and Co layers and chosen Co at
the interface with MgO.  Nonetheless, even the simplified structure
can give us some insights into the impact of the type of electrode on the
defect levels.

For the calculations involving full junctions, the kinetic energy cutoff value
and the convergence criterion were the same as for the bulk MgO, but the value
of $\tau$ was decreased to 0.1 eV and the spin polarized version of PBE-GGA was
used.  To compute the electronic structure of oxygen vacancies, the lateral
directions of MTJ were doubled  and periodic boundary conditions applied in all
directions.

\subsection{Transport calculations}

The ballistic conductance is  calculated using Landauer-B\"{u}ttiker \cite{Landauer1987,Buttiker1985} formula

\begin{equation}
G(E_F)=\frac{e^2}{h}\sum_{n,\mathbf{k},\sigma} T_{n,\mathbf{k},\sigma}(E_F),
\label{eq:G_tot}
\end{equation}
where $T_{n,k,\sigma}(E_F)$ is the total transmission at the Fermi energy and the summation is over all bands $n$ crossing the $E_F$, for each
${\bf k}$ point
and spin $\sigma$. The electron transmission was evaluated using the scattering based approach with a plane wave basis set and ultrasoft pseudopotential
(USPP) scheme as implemented in the PW\textsc{cond}\cite{PWcond_2} module of the Quantum Espresso (QE)\cite{QE-2009} package.

The conductance was evaluated between two semi-infinite electrodes connected by
a scattering region that contains an insulating MgO spacer and a part of the
leads on each side of the spacer. To compute the electron transmission at a
given energy E, we first calculated the total energy of the ground state
properties  with the  PW\textsc{scf} code from QE package\cite{QE-2009} and determined
the effective potential. We then constructed the generalized Bloch states,
including propagating and evanescent states, as a solution of Kohn-Sham
equations at energy E for the infinite periodic leads, and the results were
used to construct the scattering states and compute the transmission across the
entire system.  Moreover, in the spin density functional picture, electrons of
different spin move independently in their different self-consistent
potentials\cite{PWcond_2}. Therefore in this approach spin flip events are not
included and the total transmission is the sum of the two spin channels such
that $T(E)=T_{\uparrow}(E) + T_{\downarrow}(E)$.

To find how many electrode layers should be contained in the scattering region,
we studied the changes of the electrostatic potential in the scattering region.
To ensure that the electron wave function changes smoothly at the interface
between the bulk of the electrode and the scattering region, the part of the
leads in the scattering region has to be big enough so that the changes induced
in the electrostatic potential due to the interaction with MgO are contained
entirely within the scattering region. If not, an artificial potential that
scatters the incoming electrons could be present and might affect the results.
By comparing  the total electrostatic potential for the scattering region to
the total potential of the bulk electrode, we found that from the 2nd-3rd
monolayer (ML) of Fe, the bulk electrostatic potential is restored. In order
to guarantee a proper geometrical matching between the scattering region and
the electrodes,  we used 4~ML of Fe on the left side of MgO and 5~ML of Fe on
the right side. For the defect calculations, we doubled the lateral size of the
junctions. When the antiparallel alignment of the electrodes was considered,
the size of the junction along the $z$ direction was doubled, such that the
composition of the supercell was Fe(P)/MgO/Fe(AP)/MgO/F(P).

In the ground state calculations with PWscf code, the cutoff energy values for
the plane wave basis set and the electron density were set to 40 Ry and 400 Ry,
respectively. The electronic occupations were broadened using a Gaussian
smearing technique with a smearing parameter $\tau$=0.02~Ry. The total energy
convergence threshold was set to 10$^{-8}$ Ry  and the electron density mixing
parameter to 0.1. For the ground state calculations of the ferromagnetic
alignment of the electrodes, we used a \textbf{k}-point mesh of
5$\times$5$\times$1, while for the antiferromagnetic alignment the same
\textbf{k}-point grid was slightly shifted out of the $\Gamma$ point in order
to speed up the convergence.
Since  we needed to use the same form of the pseudopotential and corresponding
exchange-correlation functional for all atoms in the junction, we chose the Perdew and
Wang (PW91) generalized gradient functional \cite{Pedrew_Wang_PW91} in a
spin-polarized form already generated and available in the QE library.

An important factor in the transmission calculations is the convergence of the
2D basis set used in the PW\textsc{cond}. Here, two parameters control the basis
set: (i) \textit{ewind} defines the energy window for reducing the 2D plane
wave basis set in the transverse $xy$ plane, and (ii) \textit{epsproj} is a
threshold for the 2D basis set reduction. The default values for the two are
$\textit{ewind} =1 Ry$  and $ \textit{epsproj}=0.001$.  Generally, the larger
\textit{ewind} and the smaller \textit{epsproj} are, the higher the accuracy of
the calculations is. However, the increase in the transmission accuracy
increases the computational cost and a suitable compromise should be found.
These parameters were tested by examining the complex band structure (CBS) of
bulk Fe and MgO\cite{PhysRevLett_95_216601,PhysicalReviewB_2006_Bowen_Dederichs_Observation}.
We found that $ewind=3,~ epsroj=10^{-6}$ were sufficient to convergence the CBS and
therefore were used also to compute the transmission. In addition, the
transmission was evaluated as a  function of the number of ${\bf k}$ points in
the 2D BZ. We tested meshes of 20$\times$20, 30$\times$30, 50$\times$50 and
80$\times$80 k-points and kept the 50$\times$50 which showed a well-converged
transmission.

\section{Electronic structure}
\subsection{MgO bulk}

Here, we briefly describe the ground state electronic properties of M center
within MgO bulk and MgO incorporated in the Fe(FeCo)/MgO MTJs. 
The removal of two neighboring oxygen atoms
from an MgO supercell results in the creation of two occupied energy levels
below $E_{\rm{F}}$. The electrons that were transferred from Mg and remain
after the oxygen removal are trapped and are mostly localized on the vacancy
sites. Since the electron distribution on the vacancies resembles that of an
oxygen O$^{2-}$ ion, the atoms around the defect are only slightly distorted.
The resulting alteration to the resulting electronic structure is very slight,
such that the distortion was neglected.

\begin{figure}[h!]
\centering
\includegraphics[width=\linewidth]{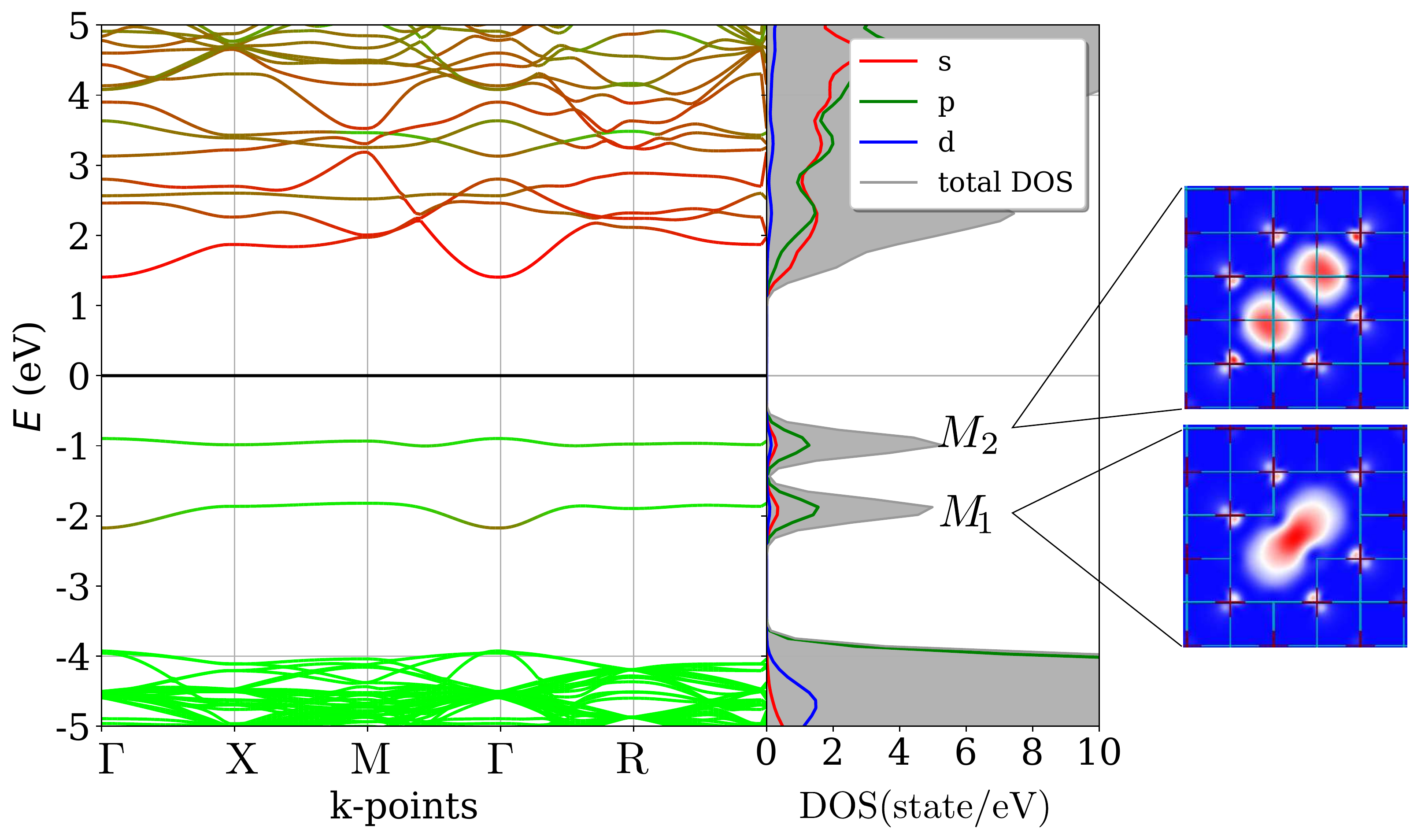}
\caption{Band structure and orbital projected  DOS for M-MgO.
Insets show the electron distribution for each of the ground state energy levels of the M center. }
\label{fig:fat_bands}
\end{figure}

To understand the nature of the M center levels, we plot in
Fig.~\ref{fig:fat_bands} the orbital-projected band structure and the density
of states (DOS) for MgO containing a M center (M-MgO). The valence states of MgO are mostly of O $p$
character while the conduction states comprise Mg $s$ and $p$-like states. The
defect levels show mostly contributions form $p$-like orbitals with a smaller
part coming from $s$-like states. By projecting the DOS on Mg and O sites, we
found that the M-levels are the results of a hybridization between O$p$
orbitals and both Mg $s$ and $p$ states. The contribution of $d$-like states is
much smaller and can be neglected. Note that the small dispersion around the
$\Gamma$ point is a result of an artificial interaction between the periodic
images of the M centers due to the 64-atom size of our supercell. As we showed
elsewhere\cite{Advanced_Materials_2017}, for supercells with 216 atoms, these
levels exhibit no energy dispersion. This means that the defects are well
separated from each other and spatially localized.  Nonetheless, the dispersion
observed for a 64-atom supercell does not significantly change the level
positions, and as such it can be neglected.

We plot as insets to Fig.~\ref{fig:fat_bands} the electronic spatial
distribution for each of the M centers ground states. It is clear that the
electron distribution for the M$_1$ state resembles a bonding-like state, while it
is anti-bonding for the M$_2$ state. As in the case of bond formation between
atoms, the coupling between two F centers creates a bonding state with a lower
energy, and an anti-bonding state with higher energy,  with respect to the
original F state energy level. Indeed, the F center peak is always positioned in between
two M center ground states. As a consequence, the barrier height created by the
F center is always higher than that associated with the M$_2$ state.

\begin{figure}[!ht]
\centering
\includegraphics[width=\linewidth]{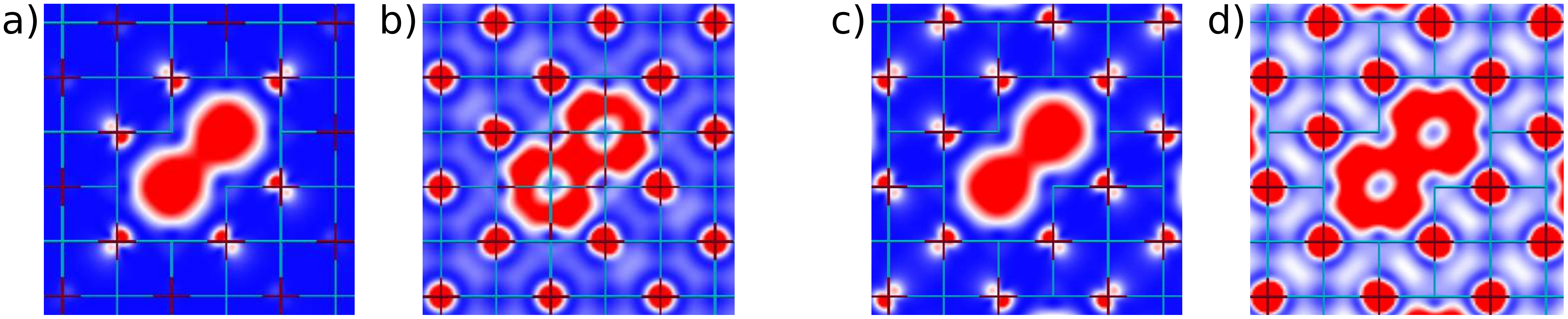}
\caption{Electron density distribution in the $xy$ plane, and within the energy
    range containing the M$_1$ and M$_2$ states (panels a/c) and M$_1^*$
    and M$_2^*$ states (panels b/d), for a 214-atom supercell (panels a/b)
    and a 62-atom supercell (panels c/d). }
\label{fgr:charge_216}
\end{figure}

Figure \ref{fgr:charge_216} presents the spatial distribution of the electron
density for both the ground (panels a/c) and excited (panels b/d) states of
M-MgO. In panels a/b (c/d), a 214-atom (62-atom) supercell was used. We observe
how neighboring oxygen vacancies hybridize to create an M center.  As expected
from the band structure plots, the electrons remaining after oxygen removal are
localized on the vacancy sites and the electrons are distributed among the
vacancies.  Since the M center's excited state lies within the conduction band
states, a nonzero electron density is present on atoms far from the defect.
The electron density plots also reveal a hybridization between the M center
states and the nearest oxygen ions, thereby showing that the defect level's excited state is
indeed mostly of oxygen $p$ character.  The spatial electron density of the M
center is fully isolated from the periodic images in the 214-atom supercell.
Thus, the lateral extent of the M center spans 1~ML on either side of the
oxygen vacancy sites that define the M center. On the other hand, spatial
overlap develops between the electron density of M centers in the 62-atom
supercell calculation. This means that M centers separated by 2~ML of MgO will
experience electronic interactions.  The electron distribution also indicates
that the ground states of the M center are mostly \textit{s}-like, while the
excited states are of \textit{p}-like character.

\begin{figure}[!ht]
\centering
\includegraphics[width=0.8\linewidth]{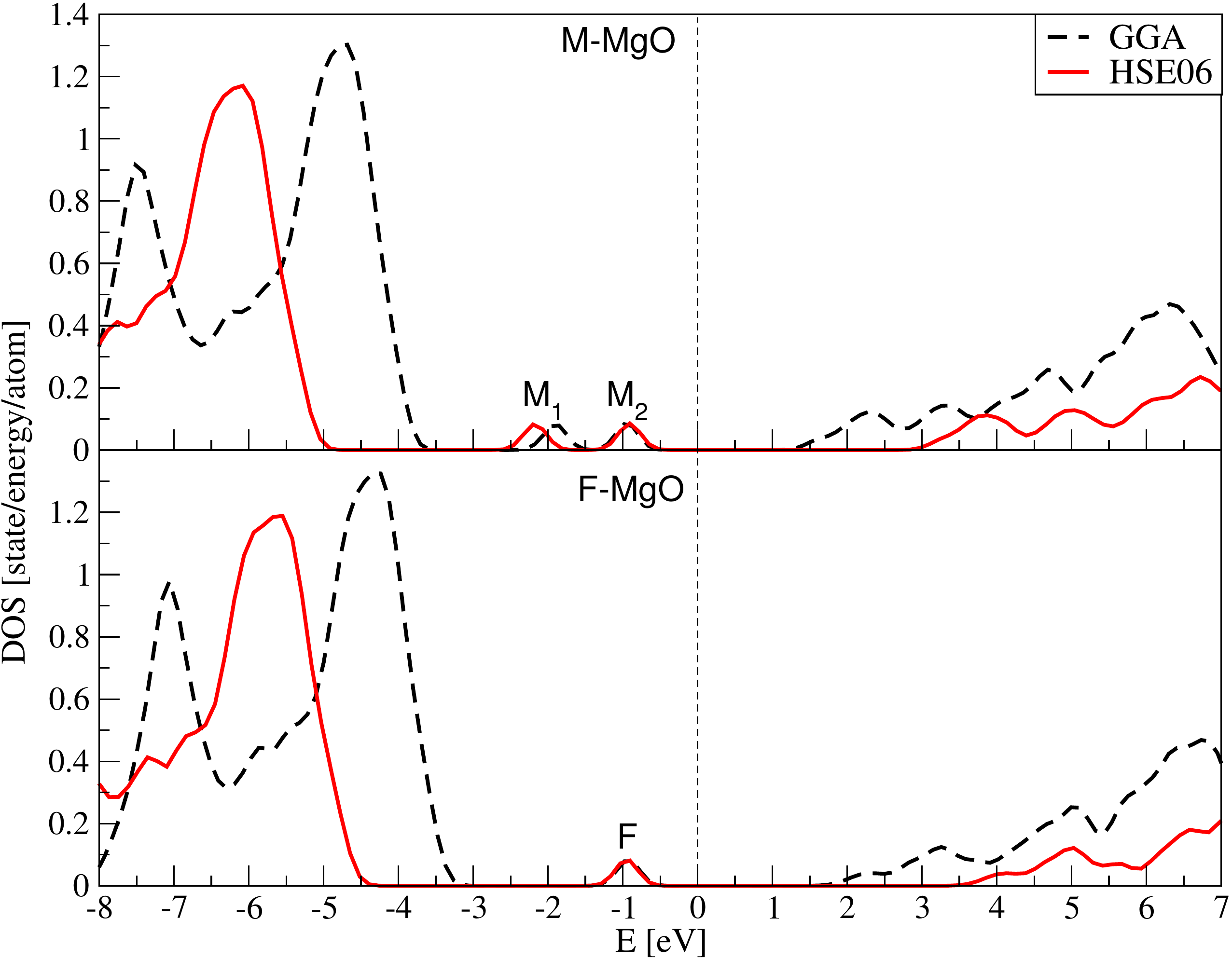}
\caption{GGA and HSE06  calculated DOS for bulk \rm{M-MgO} (top panel) and \rm{F-MgO}
    (bottom panel) with E$_F$ aligned at the zero of energy for both functionals.}
\label{fgr:Fig_1}
\end{figure}

The results presented above are obtained within the GGA functional, which is
known to underestimate the band gap. Therefore, as discussed in the methodology
section, we also employed the HSE06 hybrid functional to accurately determine
the defect level positions within the MgO band gap.  Figure~\ref{fgr:Fig_1}
presents the results of GGA and HSE06 bulk calculations for both M-MgO and
MgO containing a F center (F-MgO), with $E_{\rm{F}} \equiv$ 0~eV. For both F-MgO and M-MgO, compared to the
GGA results, the hybrid functional causes a shifting of the valence and the
conduction bands towards lower and higher energies, respectively.  The hybrid
functional, due to the inclusion of a portion of the exact Fock exchange which
is orbital dependent, increases the localization by reducing the
self-interaction error appearing in GGA. This fact has almost no influence on
the F state position since it is a single localized level.  However, in the
case of a M center, where two additional energy levels are created in the MgO
band gap, the difference can be noticeable and we observe a slight shift of
the M$_1$ state further away from the M$_2$ level. This difference of the $M_1$
position between the GGA and the HSE06 calculations is about 0.27~eV.
Nonetheless, aside from the shift in the M$_1$ energy position, we otherwise
obtain a similar energy dependence of the DOS. This shows how less intensive
GGA-based calculations already yield a qualitatively correct picture of the
electronic properties of oxygen vacancies in MgO.  It is worth noticing that
the defect levels are placed near the middle of the MgO band gap irrespective
of the type of functional used.

\subsection{M-MgO/Fe(FeCo) junctions}

\begin{figure}[!ht]
\centering
\includegraphics[width=\linewidth]{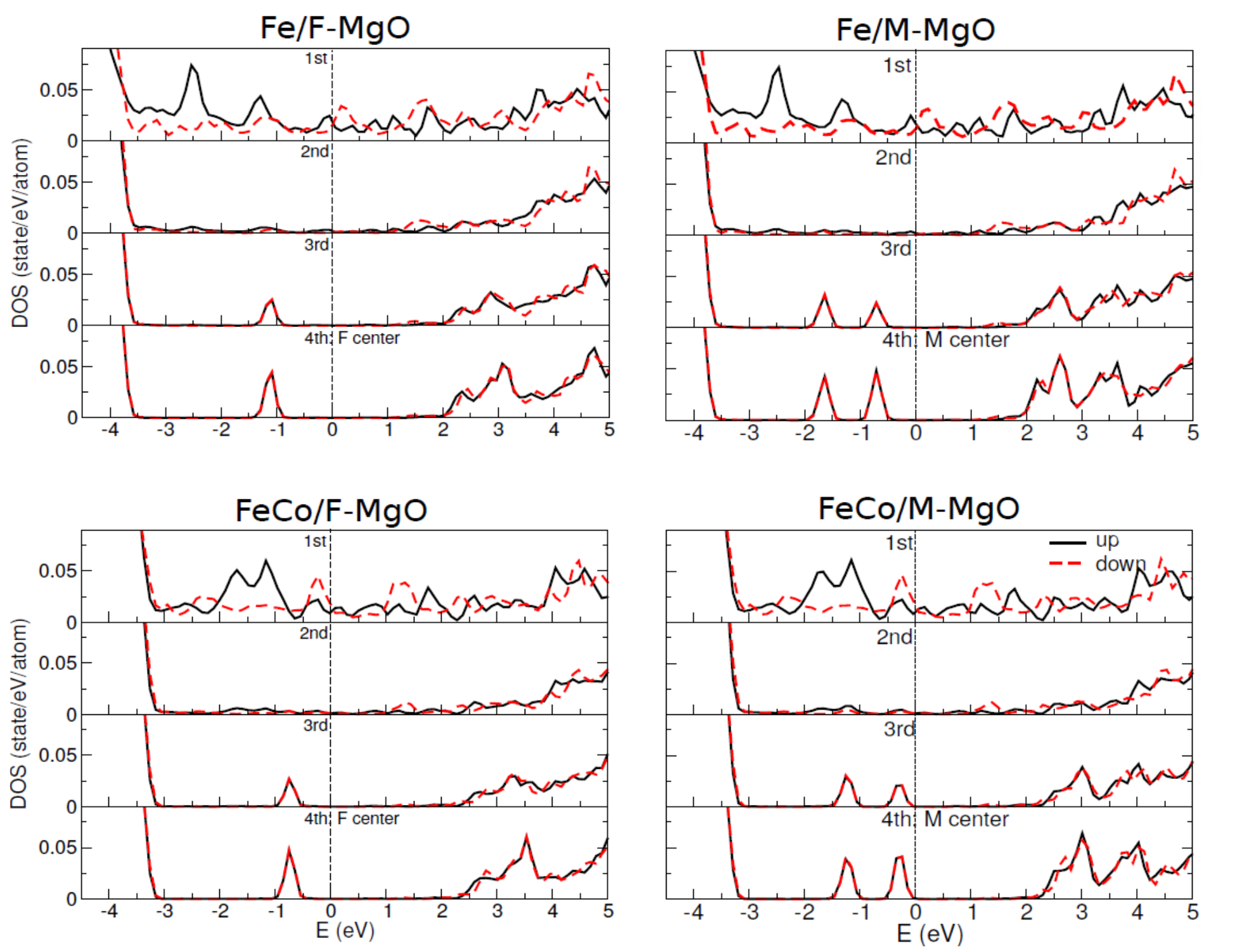}
\caption{Spin-polarized layer-projected DOS for Fe/F-MgO/Fe, Fe/M-MgO/Fe,
    FeCo/F-MgO/FeCo and FeCo/M-MgO/FeCo system with the F and M centers placed on
    the 4th layer.}
\label{fgr:Fe_Mxy}
\end{figure}

We now calculate the electronic structure of both Fe/MgO/Fe and FeCo/MgO/FeCo junctions with oxygen vacancies in the MgO spacer layer.
Fig.~\ref{fgr:Fe_Mxy} presents the DOS projected  on MgO(7ML) layers with F/M
center generated in the middle 4th layer of the MgO spacer. Due to the contact
with the metallic electrode, MIGS appear in the MgO band
gap and decay with the number of MgO layers. As a result the band gap of MgO
disappears at the interface because of states coming from Fe. From the third
layer the band gap of a bulk MgO is restored. Moreover, the ferromagnetic
electrode induces spin polarization in the neighboring MgO layers and the
difference in the DOS of spin up and down electrons can be clearly seen.

For the Fe/MgO heterostructures, the  M$_1$ and M$_2$ states are located
  respectively at -1.7~eV and -0.7~eV below the Fermi level, while the F center   level is at -1.2~eV.
  As expected the barrier height associated with the F   center is in between the M$_2$ and M$_1$ levels. 
  In the case of the FeCo   electrodes the defect levels are shifted towards the Fermi level by about   0.5~eV.
  It is worth noticing that not only are defects levels shifted, but
  also the whole band structure of MgO is rigidly shifted towards higher
  energies. This shift can be understood considering the 0.5~eV difference in
  the work function between Fe and Co. As before, the vacancy affects also
  the closest MgO layers up to 3~ML of MgO along the direction perpendicular
  to the M-center plane.

We also studied the effect of shifting the M center within the MgO~(7~ML)
spacer and found that the defect energy level remains practically unchanged
when the vacancy approaches the interface (data not shown)\cite{Taudul_Thesis}. However, if
the vacancy is placed on the interfacial MgO layer the DOS associated with the
F/M center is washed out due to the strong interaction with the ferromagnetic
electrode.  Clearly, when the defect is closer to the interface, the
hybridization between the two types of materials is strongly affected, which in
turn influences the position of the Fermi level. However, it is surprising that
this modification led only to a small differences in the Fermi level positions
of the order of 0.09~eV.

Thus far, we have considered only the situation where the M center is in the
plane parallel to the interfaces. In that case, we found that the effective
size of the M center in the direction perpendicular to the interface reaches up
to 3 ML of MgO due to charge transfer onto adjacent MLs. We can also rotate the
M center such that it be partially aligned along the $z$ direction and shared
between two neighboring MgO layers. In that case, the effective M center size
reaches 4ML. Regardless of the orientation of the defect plane, the level
position of the M center remains practically unchanged and similar results for
the DOS are also obtained (not shown)\cite{Taudul_Thesis}. Even though the changes in the DOS upon
shifting or rotating the M center within the MgO spacer are not significant, we
will show later that these changes have a huge impact on the
transmission as hinted by complex band structure
calculations\citep{Advanced_Materials_2017}. Thus, based on our calculations,
we can state that the computed energy range for barrier heights associated with
an M center can be associated with those measured experimentally at 0.4~eV, due to the M$_2$ state located $\approx$0.4~eV below the Fermi level for a FeCo/MgO/FeCo MTJ.

\section{Ballistic transport}

\subsection{Ideal MgO-based junctions}

We first calculated the transmission at the Fermi level for junctions with an
ideal, 5~ML-thick MgO structure in order to examine the impact of introducing oxygen
vacancies.  The results found for the 7~ML spacer are similar to these for 5~ML
and will be only briefly discussed.

Fig.~\ref{T_kxky_ideal_Fe} presents the transmission in the two dimensional
Brillouin Zone (2D BZ)  for the  parallel electrode magnetization for the spin
up and the spin  down electron channels (left and  middle panels) and the
corresponding transmission for the antiparallel configuration (right panel). In
agreement with previous theoretical predictions \cite{Mathon,Butler}, we found
that the majority electron transmission is centered around the $\Gamma$ point
and dominated by the $\Delta_1$ symmetry. The transmission for the minority
channel occurs  basically at the edges of the 2D BZ and is much smaller than
for the majority channel. The transmission in the AP configuration is a
mixture of features seen in both spin channels.

\begin{figure}[!htb]
     \begin{tabular}{ccc}
         \includegraphics[scale=0.4]{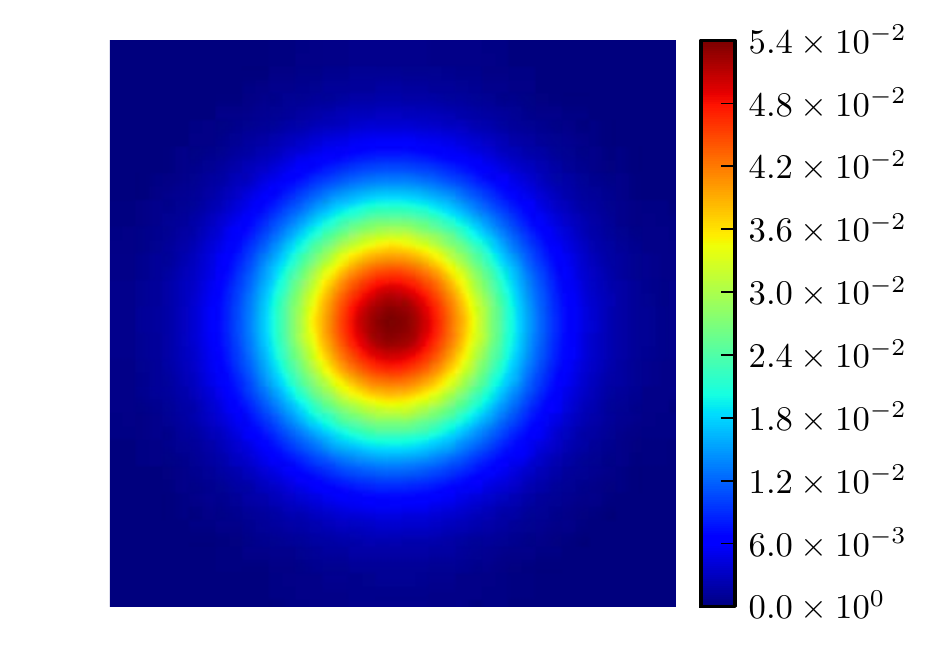} &
                \includegraphics[scale=0.4]{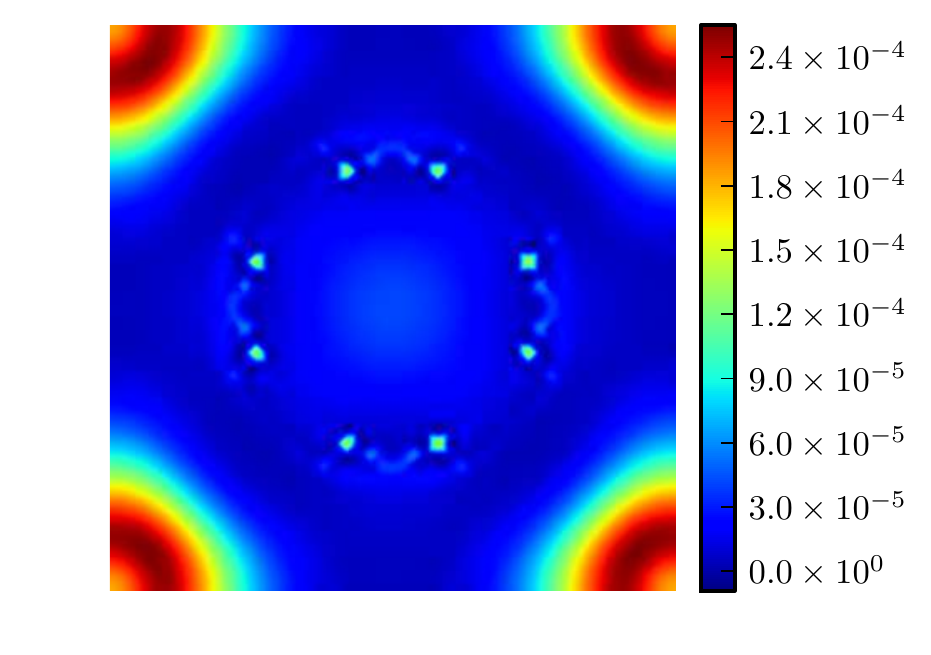} &
                \includegraphics[scale=0.4]{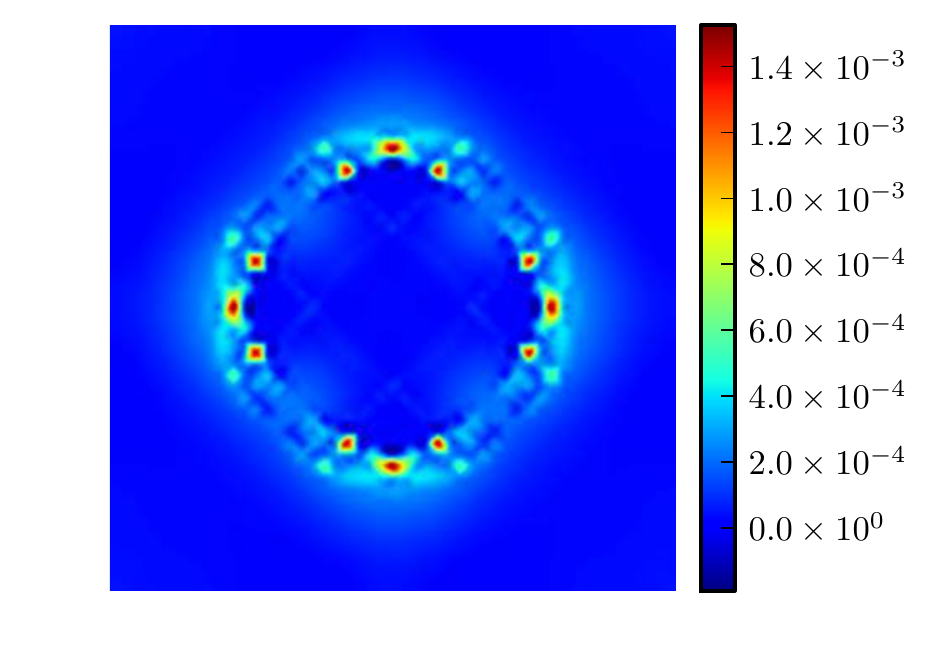} \\
         (a) & (b) & (c)
 \end{tabular}

\caption{Transmission in the two dimensional Brillouin zone (2D BZ) across an
    ideal Fe/MgO(5ML)/Fe junction in its P magnetic state for (a) the spin up
    channel, (b) the spin down channel, and (c) in its AP magnetic state.  }
\label{T_kxky_ideal_Fe}
\end{figure}

By summing the transmission over the BZ for each channel, we obtain the
conductance (see eq.~\ref{eq:G_tot}) and the resulting TMR.
Tab.~\ref{total_transmission_F_M_5ML} summarizes the transmission results for
junctions with 5 and 7~ML of MgO. As expected, the transmission decays
exponentially with the thickness of the MgO spacer and hence drops by at least
one order of magnitude when passing from 5 to 7~ML of MgO and the TMR increases
with the number of MgO layers. This reflects the favorable symmetry filtering
across MgO(001) of $\Delta_1$ electrons with a high spin polarization at the
bcc(001) Fe electrodes's Fermi level. For this reason, the $\Delta_1$ channel
is blocked in the MTJ's AP magnetic state, such that transmission is ensuring
by $\Delta_5$ and $\Delta_{2'}$ symmetry channels, which appear for both spin
populations at $E_F$ in Fe.  When the MgO thickness is increased, the
contributions to the conductance from the strongly attenuated $\Delta_5$ and
$\Delta_{2'}$ channels become smaller. This leads to a bigger overall
difference in the transmission between the P and the AP configurations and
causes  the increase in TMR. According to literature\cite{Ke2010}, the TMR
value should continue to grow up to 13~ML of MgO due to the the dominant
$\Delta_1$ contribution.  After exceeding this thickness, the TMR will also
start to decrease due to the exponential decay of the tunnelling current.

\subsection{F/M center in the middle layer of MgO}

\begin{figure}[!htb]
\includegraphics[width=\linewidth]{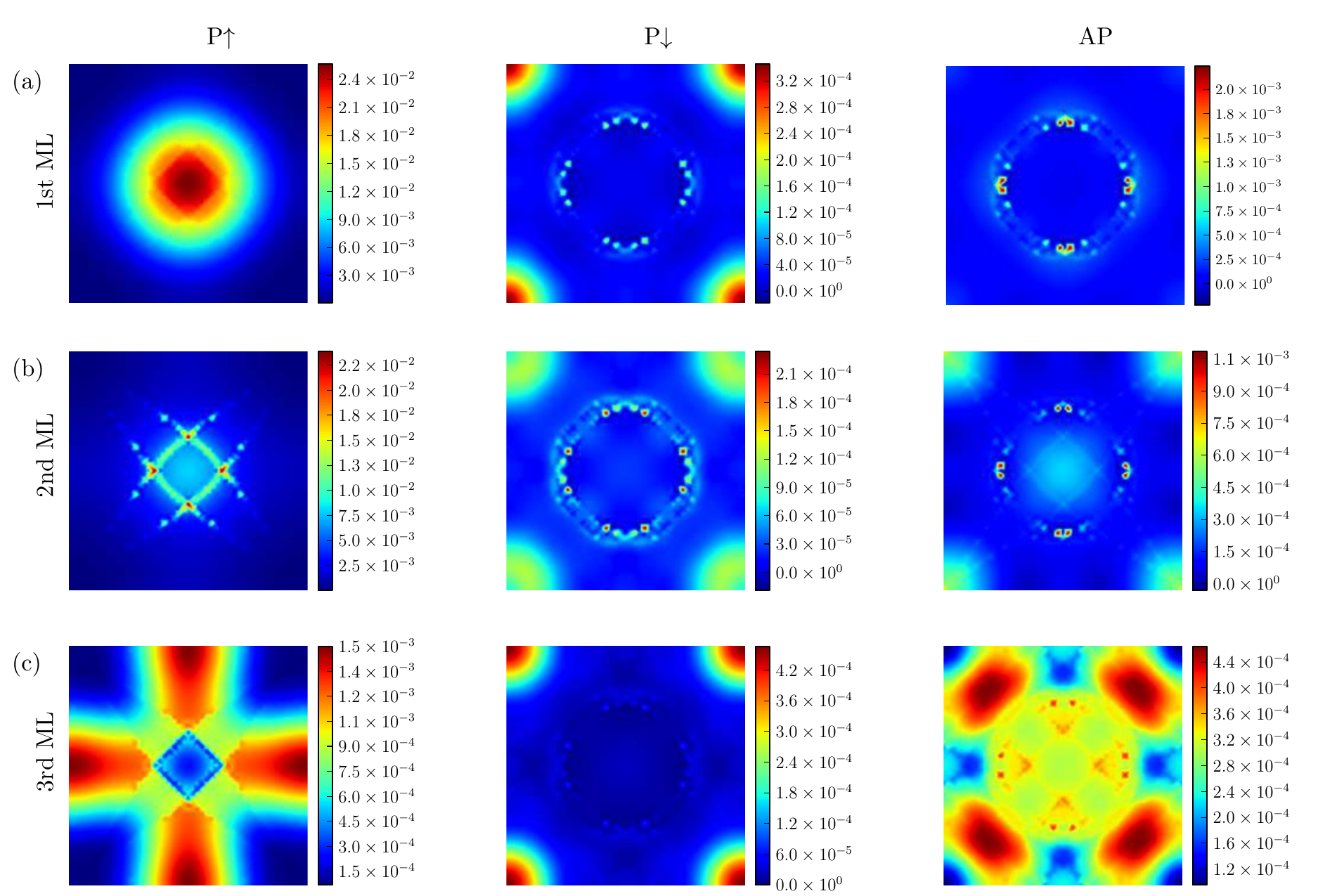} \\
\caption{
    Parallel alignment spin up (left) spin down (middle) and antiparallel (right)
transmissions in the 2D BZ for Fe/F-MgO/Fe junction for 5ML of MgO with F center in a) first ML (TMR=4261\%), b) second ML (TMR=1239\%) and c) third ML (TMR=145\%).}
                \label{fig:T_kxky_all_F_centers}
\end{figure}

\begin{figure}[!htb]
    \includegraphics[width=\linewidth]{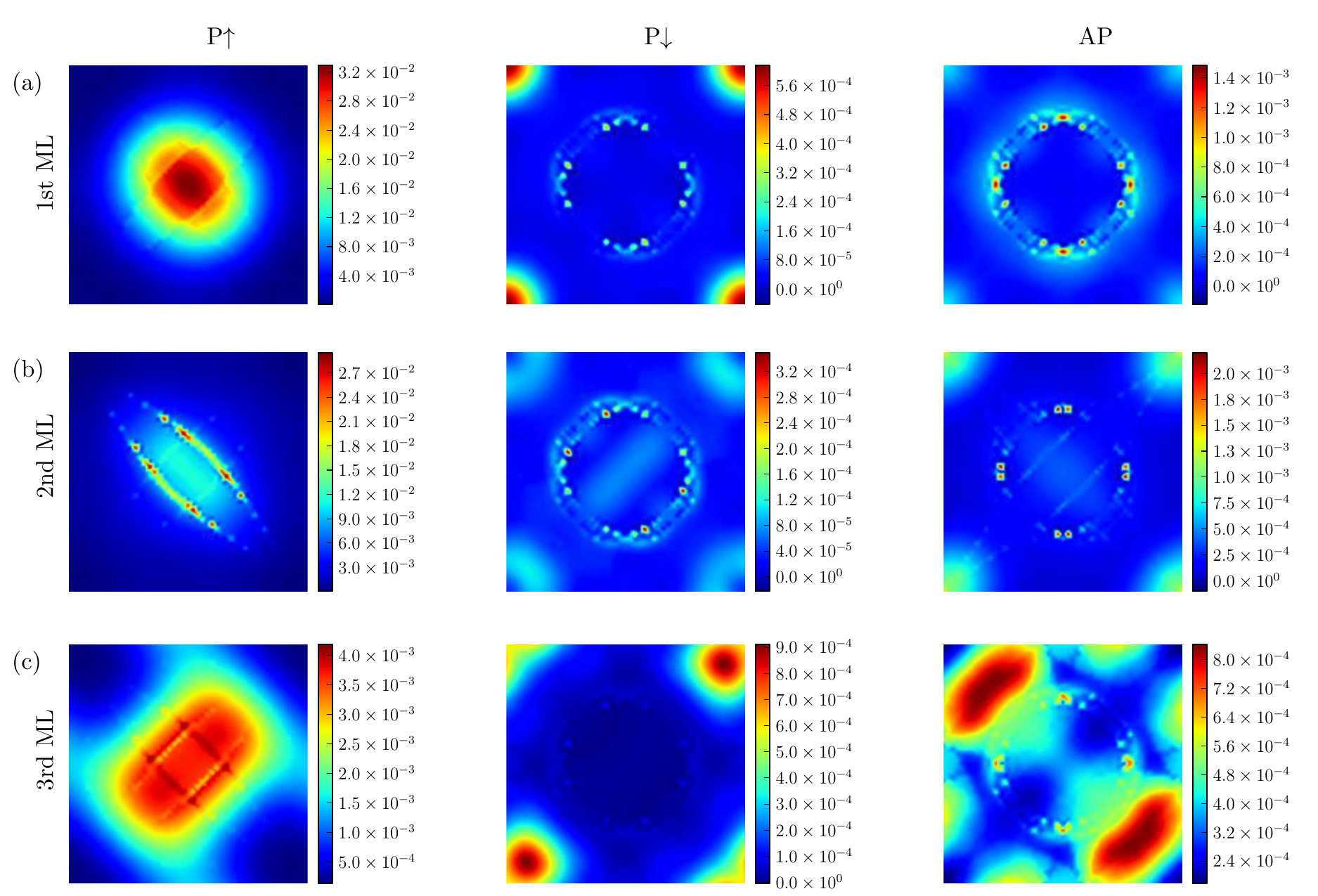} \\
\caption{Parallel alignment spin up (left) spin down (middle) and antiparallel (right) transmissions in the 2D BZ for Fe/M-MgO/Fe junction for 5ML of MgO with M center in a) first ML (TMR=3911\%), b) second ML (TMR=1135\%) and c) third ML (TMR=315\%).}
\label{fig:T_kxky_all_Mxy_centers}
\end{figure}

In the next step, we introduced single and double oxygen vacancies in the middle layer of the MgO
spacer. First, the M center's two oxygen vacancies were placed within the middle
layer. Fig.~\ref{fig:T_kxky_all_F_centers}(c) and
\ref{fig:T_kxky_all_Mxy_centers}(c) show the corresponding 2D BZ transmission
for F and M centers, respectively. The BZ transmission distribution for P spin
$\downarrow$ electrons is only slightly affected by the presence of the
vacancies, resulting in a small increase in the total amplitude with respect to
the ideal case (see Table~\ref{total_transmission_F_M_5ML}).  However, in the
spin $\uparrow$ channel, a clear
distinction  in the P transmission between the F and the M centers can be made. It appears that the F
center scatters the propagating electrons to states with higher
$\bfk$-vectors. As a result, the transmission has a minimum at the $\Gamma$
point and occurs mostly along $k_x$ and $k_y$ directions with maxima at the
edges of the 2D BZ. The electrons are scattered symmetrically in each direction
due to spherical symmetry of a single oxygen vacancy.  On the other hand, the
P spin up transmission in the presence of the M center becomes broadened in the 2D
BZ but maintains a symmetric maximum at the vicinity of   the $\Gamma$ point. This
clearly suggests that, while the transmission across a F center is reduced by
an order of magnitude due to transport across k$\neq 0$ states, coherent
transport that preserves spin and symmetry of the electron wave function is
still possible when M center is present in the MgO spacer.

We find that, while F and M centers promote a reduction in total Pup
transmission (see Tab.~\ref{total_transmission_F_M_5ML}), both centers promote
an increase in both the Pdn and AP total transmissions. Furthermore, the
transmission distribution in the AP configuration changes significantly from
that of an ideal MgO junction (see Fig.~\ref{T_kxky_ideal_Fe}). While introducing
defects reduces the TMR, the TMR is higher for M centers compared to F centers.
We found similar trends for the 7~ML , (see
Tab.~\ref{total_transmission_F_M_5ML}). Again, if we increase the number of MgO
layers (here from from 5 to 7~ML), the TMR also increases regardless of the defect type.

\begin{table}[!ht]
\begin{tabular}{l  c c c c}
 & P-UP  & P-DOWN &  AP  & TMR [\%] \\
    \hline
    \hline
 5 ML & 79.0  & 0.46 & 1.0  & 7850 \\
F (5ML)  & 7.21 & 0.63 & 3.2 & 145 \\
M (5ML) & 17.1 &  1.47 & 4.5 & 315  \\
    \hline
7 ML & 5.3          &0.003 & 0.03 & 15770   \\
F (7ML) & 0.12      & 0.006 & 0.03 & 304 \\
M (7ML) & 0.62      &  0.007 & 0.04 & 1624 \\
    \hline
    \hline
\end{tabular}

\caption{Total spin polarized transmissions $\times 10^4$  and  TMR for Fe/MgO/Fe, Fe/F-MgO/Fe
and Fe/M-MgO/Fe junctions, each with 5 and 7ML of MgO. The F/M center is located in the middle layer. }
\label{total_transmission_F_M_5ML}
\end{table}

Since the M center promotes a 0.4eV barrier height in MgO MTJs with FeCo electrodes,
these transmission results can account for the simultaneous experimental occurrence
of high TMR alongside 0.4eV barrier heights. They also confirm the initial assumption
that coherent transport can be preserved when a M center is present.
Note that the defect level positions discussed previously were evaluated using
the VASP code with the PAW basis set. To verify the robustness of these
results, we switched to a plane wave basis set in conjunction with an ultrasoft
pseudopotential approach.  While the shape of the layer-projected DOS is
practically the same, we noticed a small shift of about 0.15~eV of the F and
M$_1$ states towards lower energies. We then examined how this shift can
influence the conductance by examining the transmission in the energy window
$E_F\pm 0.1$ eV.  In the case of the spin up transmission,
the 2D~BZ distribution and the amplitude of the transmission for all structures
remains practically the same.  However, some changes were observed in the spin
down transmission. The likely cause is the presence of minority interfacial
resonant states (IRS)\cite{Butler_2008}. This discrepancy should not influence the
generality of the results presented since the contributions from the spin down
channel to the $G_{\rm P}$ are much smaller than those of the spin up channel.

\subsection{Effect of shifting the vacancy on the transmission}

We now examine the impact on spin-polarized transmission of varying the position within the barrier of the F and M centers alters. Although we did not observe any significant change in the layer-projected DOS upon moving the vacancy to the interface, the transmission was nevertheless profoundly altered.

Fig.~\ref{fig:T_kxky_all_F_centers}(a) and \ref{fig:T_kxky_all_Mxy_centers}(a) respectively show the 2D BZ spin-dependent transmissions with F and M centers generated in the interfacial MgO layer. Here, the M center always remains in the plane parallel to the interfaces.  Interestingly, we found that the transmission distribution is almost the same as for the ideal junction with the peaks amplitude very close to the ideal case (compare with Fig.~\ref{T_kxky_ideal_Fe}). The calculated TMR reaches about 4261\% and 3911\% for the F and M center, respectively, i.e. are of the same order of magnitude as for the ideal junction (see Tab.~\ref{total_transmission_shifting}).

\begin{table}
\centering
\caption{Total spin polarized transmission$\times 10^{4}$
    and TMR for Fe/F-MgO(5ML)/Fe and Fe/M-MgO(5ML)/Fe junctions with vacancies shifted within MgO layers.
    }
\begin{tabular}{l   c c c c}
             & P-UP  & P-DOWN &  AP  & TMR [\%] \\ \hline
F in 1st ML  & 54.5 & 0.4 & 1.26  & 4261  \\
F in 2nd ML  & 20.3 & 0.4  & 1.55 &  1239 \\
F in 3nd ML  & 7.21 & 0.63  & 3.2 &  145 \\ \hline
M in 1st ML & 67.6 & 0.6 & 1.70  & 3911 \\
    M in 2nd ML & 30.3 &  0.5 & 2.49 & 1135 \\
    M in 3nd ML & 17.1 &  1.47 & 4.5 & 315
\end{tabular}
\label{total_transmission_shifting}
\end{table}

When vacancies are on the second layer from the interface, the transmission
decreases and we observe additional sharp spikes in the P spin up channel
(Figs.~\ref{fig:T_kxky_all_F_centers}(b) and
\ref{fig:T_kxky_all_Mxy_centers}(b)).  The P spin down and AP transmission
distributions are only slightly affected.  Note that the layer alternation also
causes a rotation of the M center within the $xy$ plane when we go from one layer to
the next one. This explains the observed rotation in the transmission amplitude
in the 2D BZ (compare for example panel (c) and (d) in
Fig.~\ref{fig:T_kxky_all_Mxy_centers}).

\begin{figure}[!htb]
                \centering
\includegraphics[scale=0.20]{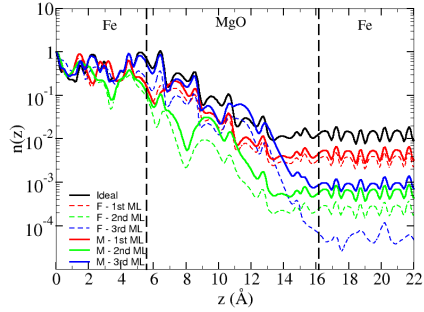}
\caption{Spin up $\Delta_1$ scattering state distribution along the transport direction $z$ for various vacancy
configurations of Fe/F-MgO(5ML)/Fe and Fe/M-MgO(5ML)/Fe  in the MTJ's P magnetic state. 
All the data are normalized and the same logarithmic scale is used. }
                \label{fig:scatteringZ}
\end{figure}

To understand the changes in the transmission when varying the F/M centers
position, we analyzed a real space distribution of scattering states at the
$\Gamma$ point. We discuss here the MTJ's P magnetic state, focusing on the
spin up channel since its transmission strongly drives the ensuing spintronic
performance. 
In the spin up
channel we focus on the $\Delta_1$ symmetry since is has the smallest
attenuation rate within MgO barrier and the biggest impact on the resulting
transmission. 
Fig.~\ref{fig:scatteringZ} shows the density of a $\Delta_1$ scattering state, summed over the $xy$
plane, as a function of the position~$z$ along the transport direction for
various defect configurations. Clearly, in the presence of vacancies, the
amplitude of the $\Delta_1$ channel is decreased with respect to the ideal
case.  Interestingly, M centers systematically yield a higher transmission
amplitude than F centers at all defect positions within the barrier.

\begin{figure}[htp]
    \begin{tabular} {cc}
        \includegraphics[scale=0.20]{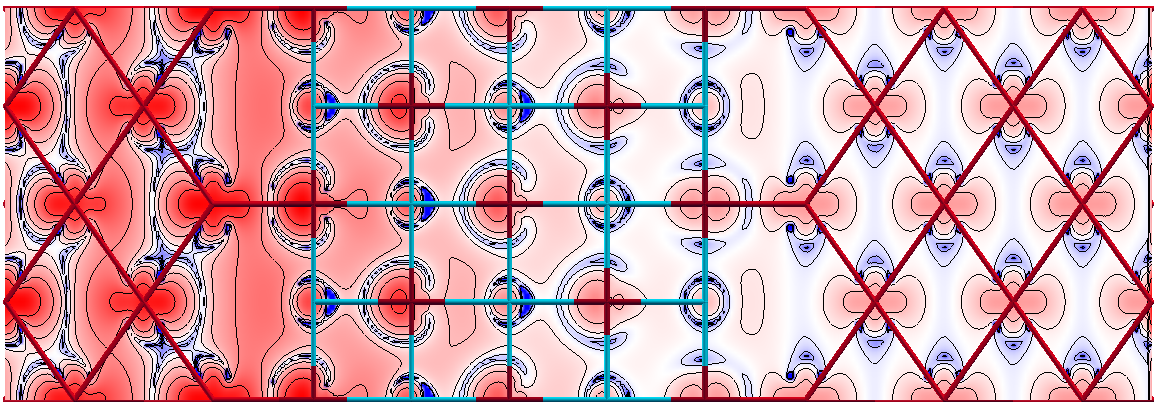} & \\
    (a) Ideal MTJ &   \\
        \includegraphics[scale=0.20]{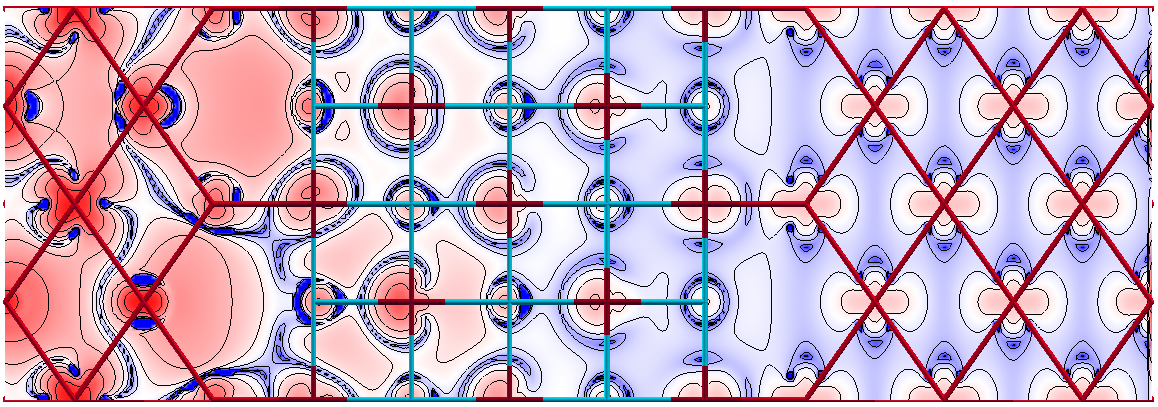} &
            \includegraphics[scale=0.20]{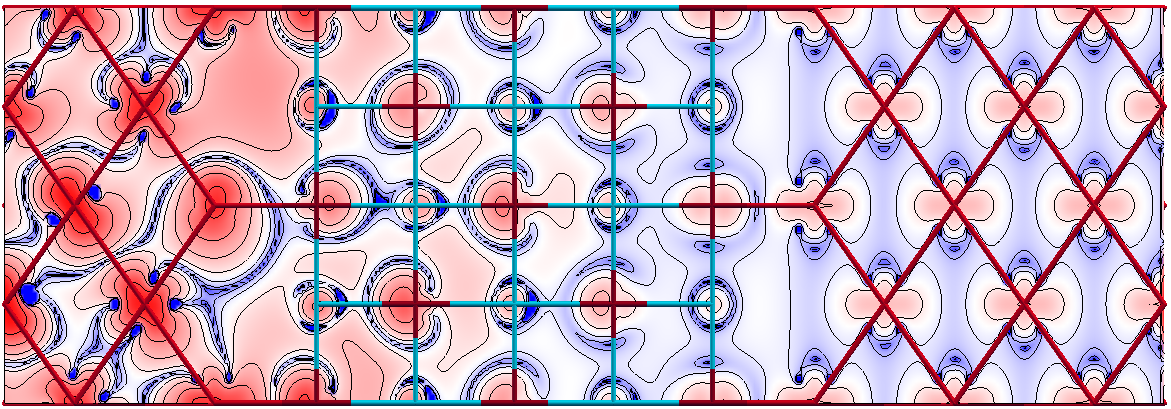} \\
    (b)    F in 1st ML & (c) M in 1st ML   \\
        \includegraphics[scale=0.20]{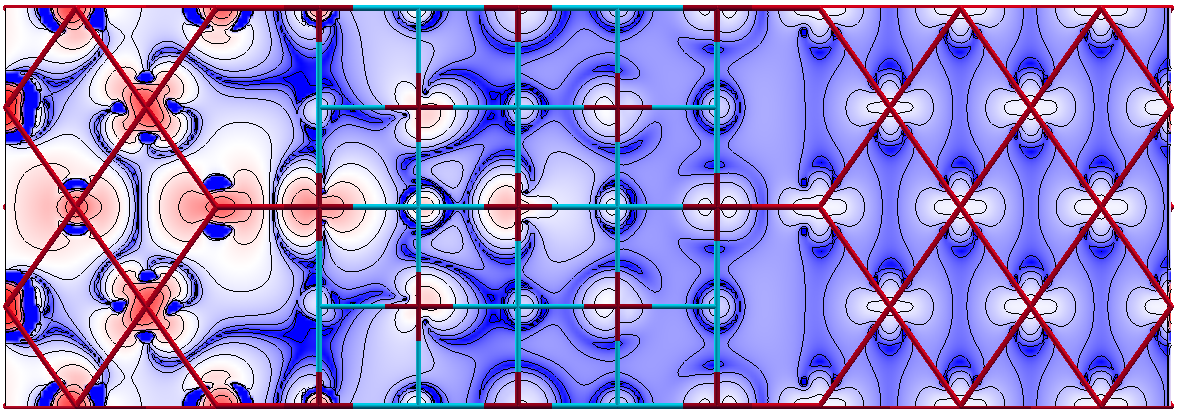} &
        \includegraphics[scale=0.20]{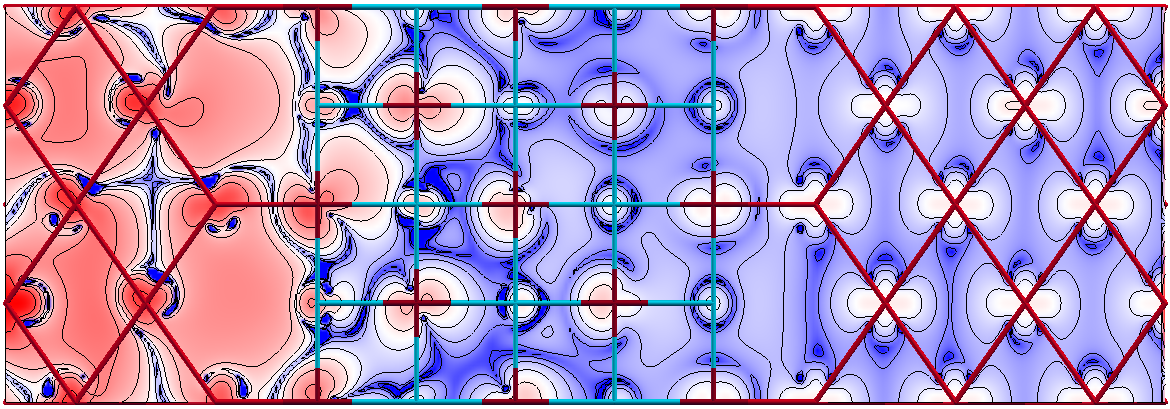} \\
    (d) F in 2nd ML  & (e) M in 2nd ML \\
        \includegraphics[scale=0.20]{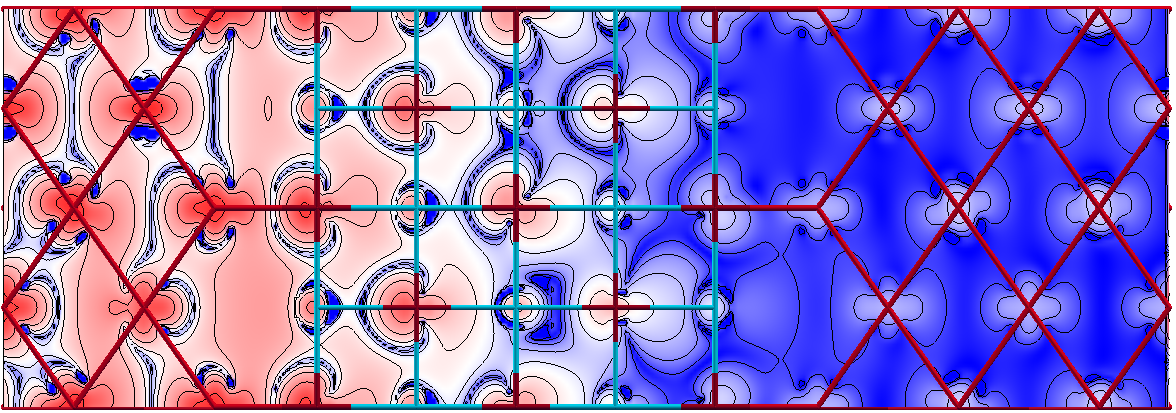} &
        \includegraphics[scale=0.20]{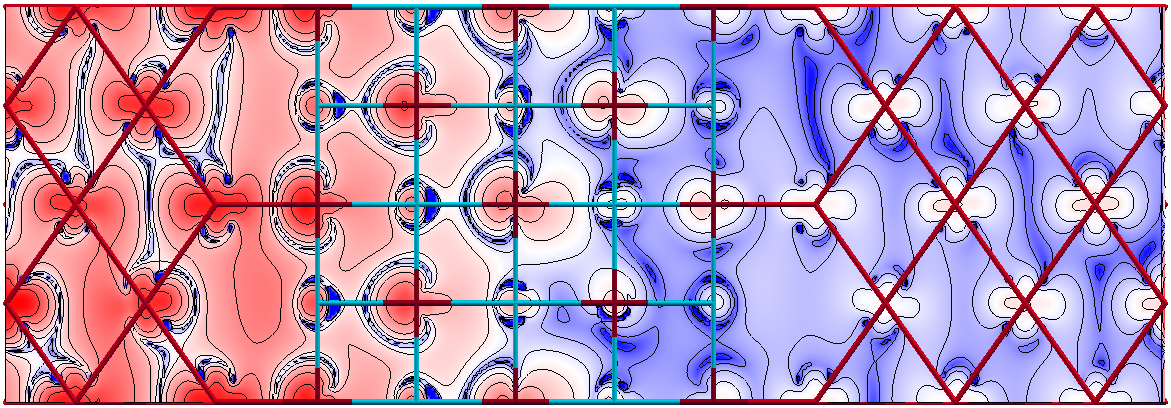}  \\
    (f) F in 3rd ML   & (g) M in 3rd ML    \\
\end{tabular}
    \caption{2D representation of the spin up $\Delta_1$ scattering states in the $x,z$ plane for
    various vacancy configurations of . All the data are normalized and the same logarithmic scale is used (see Fig. \ref{fig:scatteringZ}.
    }
    \label{fig:scattering_states_UP}
\end{figure}

Fig.~\ref{fig:scattering_states_UP} shows the $\Delta_1$ scattering states at
the Fermi level across a Fe/MgO(5ML)/Fe junction for the ideal case and the
various positions of the F and M centers. 
All the data are normalized and the same logarithmic scale is used for comparison purposes.

As expected, in the case of an ideal junction, the $\Delta_1$ channel originates from the left
electrode, crosses the MgO barrier and ends in the right electrode. When
F/M centers are introduced, the distribution of the $\Delta_1$ state changes
and depends on the vacancy type and position. The most beneficial configuration
is with the vacancies located at the interfacial MgO layer. In that case, the
amplitude of the scattering stated is just slightly lowered with respect to the
ideal situation, and the $\Delta_1$ channel is still transmitted from the left
to the right electrode. Moreover, these graphs indicate that the further from
the interface a vacancy is, the bigger the difference between F and M centers.
The difference in distribution of the $\Delta_1$ channel for F and M center in
the third layer of MgO (Fig.~\ref{fig:scattering_states_UP}) can explain the
resulting values of TMR, 145\% and 304\% respectively, which underscores the
synergistic spintronic role of M centers compared to F centers.


The overall picture is that the transmission of the spintronically crucial
$\Delta_1$ spin $\uparrow$ channel in the MTJ's P magnetic state in the presence of
vacancies 1) is higher in presence of an M center rather than an F center for a defect positioned at the center of the barrier, and
2) is close to that of an ideal junction when either vacancy type is positioned near
the interface. This second point sheds precious light into how a MgO-class MTJ
can experimentally exhibit both high TMR and a low barrier height. Indeed, the
MgO barrier is often formed atop the FeCoB metallic surface by sputtering
metallic Mg, followed by an oxidation step\cite{JJAP_2012_Dahmani_oxidation,Dave_2006}. Avoiding the oxidation of the lower FeCoB interface can naturally lead to the presence of oxygen vacancies within
the first ML of MgO. Interfacial oxygen vacancies also play a role in promoting perpendicular magnetic anisotropy for ultrathin ferromagnetic films in MTJs\cite{PhysRevB.84.054401,RevModPhys.89.025008}. Our results show that, counterintuitively, such states can maintain near-ideal levels of TMR and promote the low barrier height needed\cite{Halisdemir2016} for spin transfer torque.

\subsection{Rotating the M center}

We now examine the impact on transmission of rotating the M center plane so
that it is shared between two adjacent MgO layers. Prior complex band structure
calculations \cite{Advanced_Materials_2017} indicate that if the M center is
located on two neighboring MgO layers parallel to the interface, the
attenuation coefficient for the $\Delta_1$ channel can be slightly smaller or
comparable to the ideal case.  To verify whether this attenuation is reflected
in the transmission, we considered a symmetric junction with 6~ML of MgO, such
that the M center is shared between the two middle layers. We also increased
the number of electrode layers included in the scattering region to ensure
proper geometrical matching at the interfaces.

\begin{figure}[!htb]
\begin{tabular}{ccc}
            \includegraphics[scale=0.5]{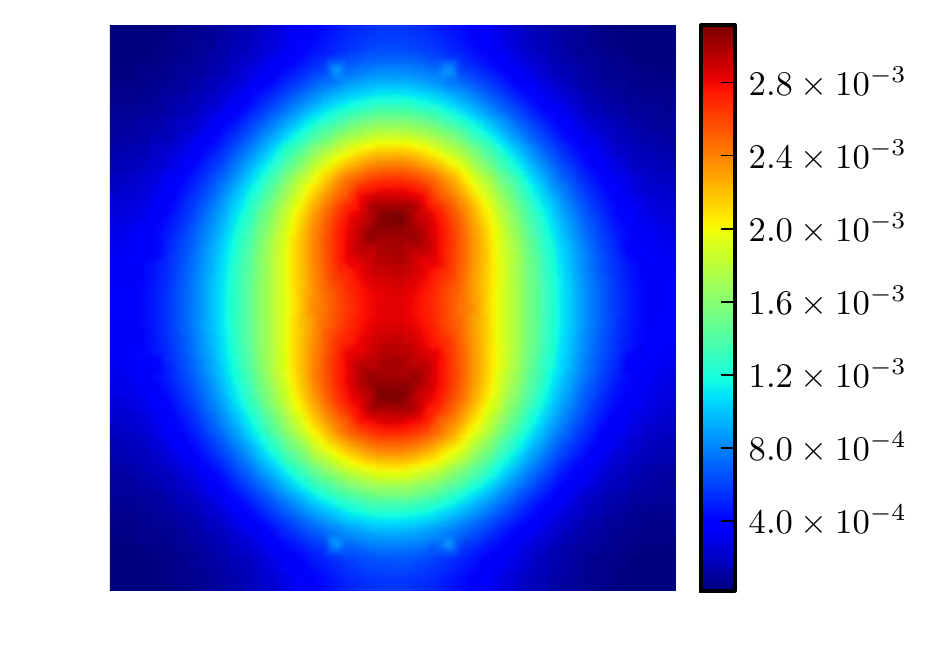} &
                \includegraphics[scale=0.5]{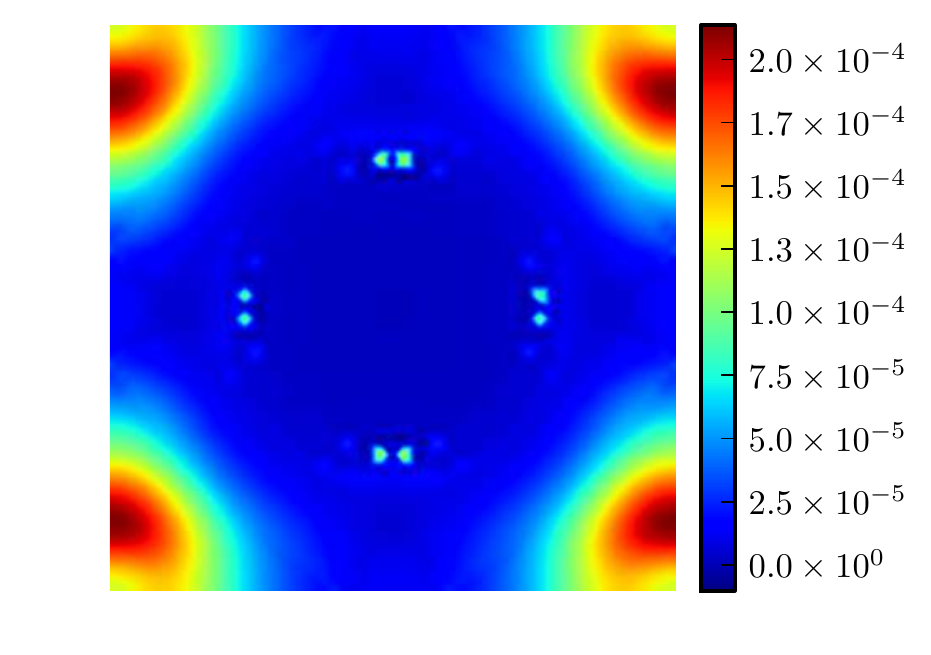} &
                \includegraphics[scale=0.5]{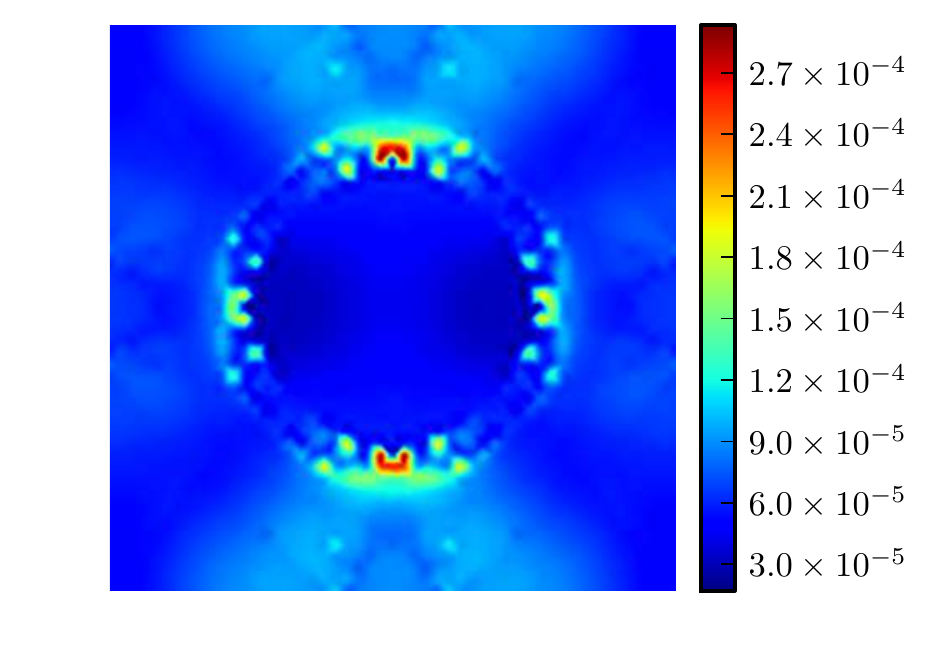}  \\
    (a) &(b) & (c)   \\
\end{tabular}
\caption{Transmission in the 2D BZ for  Fe/M-MgO/Fe junction with 6~ML MgO spacer and the M center located on the MgO 3rd and 4th ML in the $yz$ plane.}
\label{Myz_transmission}
\end{figure}

Fig.~\ref{Myz_transmission} presents the transmission in the MTJ's P magnetic state for both spin channels, and in the AP magnetic state. Comparing with the ideal case (Fig. \label{T_kxky_ideal_Fe}), we find that both the P spin down and AP transmissions are practically unaffected by the defect. The spin up transmission is even more concentrated around the $\Gamma$ point than before. The TMR value reaches 1423\%, which is as high as when the F/M center is positioned next to the interfacial layer.

If we compare the spin up transmission in Fig.~\ref{Myz_transmission} and
\ref{fig:T_kxky_all_Mxy_centers}(c), we infer that the shape of the
transmission reflects the symmetry/orientation of the M center. Indeed, when
the M center is generated in a MgO plane parallel to the interfaces, the two
oxygen vacancies lie along the diagonal, and a propagating electron
simultaneously encounters both oxygen vacancies. This explains the elongation
of the transmission peak along the diagonal of the plane in
Fig.~\ref{fig:T_kxky_all_Mxy_centers}(c). On the other hand, when the M center
is partially along the transport direction, i.e., in the $yz$ plane, such that
the two oxygen vacancies are in adjacent $xy$ planes, the propagating
electron reaches the first oxygen vacancy and then the second.  As a result,
the transmission is now along the  $k_y$ direction in the BZ
(Fig.~\ref{Myz_transmission}). The transmission peaks are of same intensity
because, owing to the symmetrical MgO spacer, electrons propagating from the
left and the right electrodes see the same potential landscape.

This preservation of high TMR thanks to a P $\uparrow$ transmission channel
that is concentrated at the $\Gamma$ point illustrates how the scenario of a M
center, at the barrier's center and partly directed along the tunnelling
direction, can also concurrently generate high spintronic performance alongside
a low barrier height.

\section{Conclusion}

We have analyzed the electronic properties of single (F centers) and paired (M
centers) oxygen vacancies in bulk MgO and in the MgO spacer of Fe/MgO/Fe MTJs,
and their impact on ballistic spin- and symmetry-polarized transport. As
detailed below, we conclude that the experimental sample preparation techniques
associated with the concurrent observation of high TMR and low barrier heights
can be theoretically explained in terms of the presence of oxygen vacancies in
the barrier,  especially near a MTJ interface.

The M center generates two doubly occupied energy levels within the MgO band
gap that mimick the bonding (M$_1$) and antibonding (M$_2$) atomic-like states
created due to two interacting F centers. As a result, the M center's
antibonding M$_2$ state generates a lower tunnelling barrier height than does
the F center. The energy level associated with a M$_2$ center is shifted from
-0.7~eV up to -0.2~eV below the Fermi level when we switch from an Fe to a Co
interface, in agreement with the 0.5~eV change in work function of the
Fe and Co surfaces. The M center's energy levels remain unchanged upon
moving the M center from the MTJ interface to the barrier middle, and upon
changing its orientation relative to the interfaces.  Our results therefore
explicitly ascribe the experimental barrier heights of 0.4~eV to the presence
of paired oxygen vacancies within the MgO barrier.

Incorporating either a F or M center within a Fe/MgO/Fe magnetic tunnel
junction can decrease the transmission of the P spin up channel, while
increasing somewhat that of the P spin down and AP channels. As a result the
theoretical TMR can drop by up to two orders of magnitude, from $~$10000\% to  $~$100\%. Overall, M centers tend to
maintain a transmission maximum at the $\Gamma$ point for the P spin up channel,
with only a small broadening, while F centers introduce scattering to higher
$\bfk$-vectors, thereby decreasing the channel's conductance. F and M centers
induce only small increases in the P spin down and AP conductances.
Consequently, the TMR is generally higher for transport across M centers than for F
centers. Since the formation energy of a M center is lower than that of two F
centers, annealing can induce the preferential presence of M centers over F
centers, which in turn promotes higher spintronic performance\cite{Advanced_Materials_2017}.

Our study indicates that the position of F and M centers crucially impacts
magnetotransport. Compared to the case of an ideal junction, defects located on
the interfacial MgO layer induce practically no change in either the shape of
the transmission distribution or its amplitude. The resulting TMR reaches
around 4000\%, and the system amounts to an ideal MTJ with a barrier of reduced
height and effective thickness. Moving the defect away from the interface
reduces the P spin up transmission, and thus TMR, especially for the F center.
This theoretical insight is compatible with the likely presence in experiments
of oxygen vacancies at the lower MTJ interface when the MgO barrier is grown by
oxidizing thin layers of metallic Mg deposited atop the lower ferromagnetic
metallic electrode while avoiding the latter's
oxidation\cite{JJAP_2012_Dahmani_oxidation,Dave_2006}. It is also in
line with the role of interfacial oxygen vacancies in promoting perpendicular
magnetic anisotropy in the adjacent ultrathin ferromagnetic
films\cite{PhysRevB.84.054401,RevModPhys.89.025008} of MgO-class MTJs with perpendicular
magnetization.  Finally, we find that orienting a M center at the barrier center so as to partly point along the transmission direction yields TMR $~$1000\%.

 Our study thus identifies conditions on the nature and positioning within the MgO of single/double oxygen vacancies so as to obtain predicted TMR values in excess of 1000\% in MTJs with low barrier heights, in line with TMR amplitudes reported experimentally\cite{ikeda_tunnel_2008}. Our theoretical results thus reconcile the simultaneous presence of high TMR and low barrier heights in MgO-class MTJs by ascribing them to the presence of oxygen vacancies. Looking ahead, the respectively 3ML and 3-4ML effective physical size of the F and M centers condition not only the minimum barrier thickness for sizeable TMR (around 3ML\cite{JournalAppliedPhysics_2010_Skowronski_Dijken_Interlayer}), but also the MTJ's lateral size. Technological progress has enabled the demonstration of working MTJs
with a lateral size down to 4.3nm\cite{Watanabe2018}.
Experiments are thus approaching the 7-8ML (i.e $\approx$2nm) limit estimated for a M
center to retain its electronic properties\cite{Advanced_Materials_2017}. Our work provides a much-needed theoretical basis to move beyond the mostly unsuspected, fortuitous defect engineering of spintronic performance that has thus far propelled MgO-based spintronics and its applications.  

\begin{acknowledgements}
This work was performed using HPC resources from the Strasbourg Mesocenter and from the GENCI-CINES Grant gem1100.
MB acknowledges funding from the Agence Nationale de la Recherche (ANR-14-CE26-0009-01)
and the Labex NIE "Symmix" project (ANR-11-LABX-0058 NIE) and   
thanks F. Schleicher, D. Lacour, M. Hehn, F. Montaigne, S. Boukari and W. Weber for useful discussions.
\end{acknowledgements}

\end{document}